\documentclass[prb]{revtex4}
\newcommand{\be}{\begin{equation}}
\newcommand{\ee}{\end{equation}}
\newcommand{\ba}{\begin{eqnarray}}
\newcommand{\ea}{\end{eqnarray}}
\usepackage{graphicx}
\usepackage{dcolumn}
\usepackage{bm}
\usepackage{amsfonts}

\begin{document}

\preprint{}

\title{Rigidity-based approach to the boson peak in amorphous solids: from sphere packing to amorphous silica }
\author{ Matthieu Wyart}

\affiliation{School of Engineering and Applied Sciences, Harvard University, 29 Oxford Street, Cambridge, MA 02138}
\date{\today}

\begin{abstract}

Glasses have a large excess of low-frequency vibrational modes in
comparison with continuous elastic body, the so-called Boson Peak, which appears to correlate with several crucial properties of glasses, such as transport or fragility.  
I review recent results \cite{matthieu1,matthieu2,these} showing that the Boson Peak is a necessary consequence of the weak connectivity of the
 solid. I explain why in assemblies repulsive spheres the boson peak shifts up to zero frequency as the pressure is lowered toward the jamming threshold, and derive the corresponding exponent.
 I show how these ideas capture the main low-frequency features of the vibrational spectrum of amorphous silica. These results extend arguments of Phillips \cite{phillips} on the presence of floppy modes in under-constrained covalent networks to glasses where the covalent network is rigid, or when interactions are purely radial. 
 
\end{abstract}
\maketitle{}
\section{Introduction}

Elasticity and transport in crystals are fairly well understood subjects, due to the simplicity  of their underlying lattice.  
By contrast, in amorphous solids disorder has strong effects at  intermediate length scales.  As a consequence energy transport,  force propagation and low-energy excitations  are not yet satisfyingly understood.  Another intriguing property is the ubiquitous presence in their vibrational spectrum of an excess of low-frequency modes with respect to the Debye prediction, the Boson Peak. 
Many observations support that this Peak is a key property of glasses: near the peak frequency  transport is strongly affected \cite{Wphillips}, the amplitude of the peak correlates well with the dynamics of the glass transition \cite{n} and with the spatial extension of the force chains \cite{wouter}.

The cause and the nature of the peak are debated questions. Various approaches have focused on the role of disorder \cite{schi,tu, parisi2,parisi3,mezard}. In the present work we will follow a more geometric and perhaps more intuitive road, based on the concept of rigidity. Following ideas of Maxwell \cite{max}, Phillips \cite{phillips} realized that as the composition of covalent glasses is changed to increase the mean valance of the atoms, the backbone of covalent bonds could undergo a rigidity transition \cite{thorpe}. Under-constrained covalent networks are floppy, and therefore present nearly zero-frequency modes (which get a small but non-zero frequency due to the presence of van der Waals interactions). Such floppy modes are indeed observed in the spectra of those weakly-connected solids \cite{kami,t}, causing a boson peak whose amplitude tends to decrease as atoms of high valence are added to the composition of the glass, as predicted. These arguments explain the nature of the  Boson Peak in under-coordinated covalent networks. It nevertheless leaves two fundamental questions unanswered: (i) why do glasses with a rigid, sufficiently connected backbone can still present a large Boson Peak, as is the case for example for silica? (ii) Can this description be applied to non-covalent glasses, known to present a Boson Peak as well?  In what follows I shall review recent results \cite{matthieu1, these} addressing these points.

One inherent difficulty in the study of amorphous solids is that the length scales  where the effects of disorder become strong is moderate. 
This lack of large parameter makes it hard to test in a stringent manner and distinguish clearly the consequences of different theories.  It was proposed by Alexander \cite{shlomon} that the situation may be different in assemblies of purely repulsive particles, such as emulsions or elastic grains, as the pressure vanishes toward zero. This occurs at a packing fraction $\phi_c$ where repulsive particles are just in contact,  called the ``jamming transition", which corresponds to the so-called ``random close packing" for mono-disperse spheres. He proposed that  $\phi_c$ is a critical point.  This idea was latter substantiated by the findings that the elastic moduli \cite{ mason, J}, the vibrational spectrum \cite{J} and force propagation \cite{wouter} display scaling behaviors near the jamming threshold. As we shall see, although the system is amorphous and isotropic, it cannot be described as a continuous elastic body on any length scale. Because the strong effects of disorder occur already at large length scales near this critical point, this model system is a lens allowing to probe the properties of amorphous solids, and the effects of disorder. 

\begin{figure}
\centering
\includegraphics[angle=0,width=7cm]{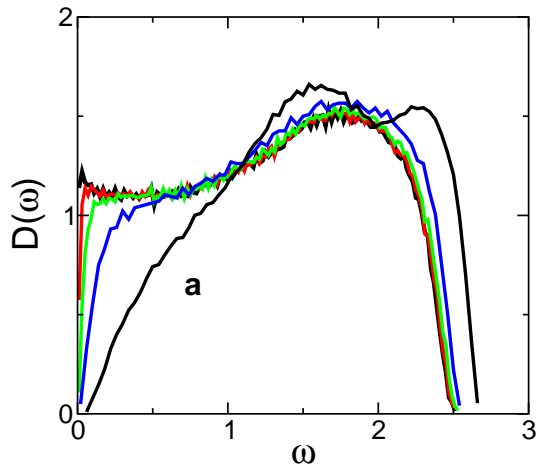}
\caption {The density of vibrational states, $D(\omega)$, vs angular frequency, $\omega$, for the simulation of Ref.~\cite{J}.  1024 spheres interacting with repulsive harmonic potentials were compressed in a periodic cubic box to packing fraction  $\phi$, slightly above the jamming threshold $\phi_c$.  Then the energy for arbitrary small displacements was calculated and the dynamical matrix inferred. The curve labeled $a$ is at a relative packing fraction $\phi - \phi_c = 0.1$.  Proceeding to the left the curves have relative volume fractions $10^{-2}$,  $10^{-3}$,  $10^{-4}$,  $10^{-8}$, respectively.  }
\label{f1}
\end{figure}

We shall focus on the vibrational spectrum, and its relation with the coordination of the packing. Numerical experiments \cite{J} showed that at the jamming threshold, the spectrum of vibrational modes $D(\omega)$ does not present a Debye behavior $D(\omega)\sim \omega^2$ on {\it any} frequency range, rather one finds that $D(\omega)\sim \omega^0$. There are no frequency-range were plane waves can be found.   As the system is compressed, this ``plateau'' erodes below some frequency scale $\omega^*$, see Fig(\ref{f1}). In what follows we shall explain these behaviors and relate $\omega^*$ to the coordination $z$ of the packing. We shall then show how these concepts apply to silica.

\section{Vibrational spectrum near the jamming threshold}

\subsection{ Energy expansion and soft modes}

Following \cite{J} we consider $N$ soft spheres packed into a
spatially periodic cubic cell of side $L$ at volume fraction $\phi$.
 In what follows we  consider repulsive, finite-range ``soft spheres''.  For inter-particle distance $r<\sigma$, the particles have non-zero mutual energy and are said to be in contact. They  interact with the following potential:
\be
\label{opo}
 V(r)=\frac{\epsilon}{\alpha} \left(1-\frac{r}{\sigma}\right)^{\alpha}
 \ee
 where $\sigma$ is the particle diameter  and $\epsilon$ a characteristic energy.  For $r>\sigma$  the potential vanishes and particles do not interact.  Henceforth we express all distances in units of $\sigma$, all energies in units of $\epsilon$, and all masses in units of the particle mass, $m$.   In the following, we consider the harmonic case $\alpha=2$, but our arguments can be applied to other potential (for example Hertzian contacts where $\alpha = 5/2$) \cite{matthieu2}.  In the harmonic case the energy expansion follows \cite{matthieu2}:
\ba
\label{1}
\delta E=  [ \frac{1}{2}\sum_{\langle  ij \rangle} (r_{ij}^{eq}-1) \frac{[(\delta\vec{R_j}-\delta\vec{R_i})^{\bot}]^2}{2 r_{ij}^{eq}} ] \nonumber
\\
+  \frac{1}{2}\sum_{\langle  ij \rangle} [(\delta\vec{R_j}-\delta\vec{R_i}).\vec n_{ij}]^2+O(\delta \vec{R}^3)
\ea
where the sum is over all $N_c$ contacts $\langle ij\rangle$, $r_{ij}^{eq}$ is the equilibrium distance between particles $i$ and $j$, $\vec n_{ij}$ is the unit vector along the direction $ij$, and $ (\delta\vec{R_j}-\delta\vec{R_i})^{\bot}$ indicates the projection of $\delta\vec{R_j}-\delta\vec{R_i}$ on the plane orthogonal to $\vec n_{ij}$. Note that Eq.~(\ref{1}) can be written at first order in matrix form, by defining the set of displacements $\delta \vec R_1 ... \delta \vec R_N$ as a $dN$-component vector
$|\delta {\bf R}\rangle$.  Then Eq.~(\ref{1}) can be written as
$\delta E = \langle\delta {\bf R}| {\cal M}|\delta {\bf R}\rangle$.  The corresponding
matrix ${\cal M}$ is known as the dynamical matrix \cite{Ashcroft}.  The $dN$
eigenvectors of the dynamical matrix are the normal vibrational modes of the particle
system, and its eigenvalues are the squared angular frequencies of these modes.

The first term in Eq.~(\ref{1}) is proportional to the contact forces. Near the jamming transition $ r_{ij}^{eq} \rightarrow 1$ so that this term becomes arbitrarily small. It has, nevertheless, interesting consequence on the mechanical stability of the system for $\phi>\phi_c$, and this term  is important to understand how the coordination $z$ varies with $\phi$, as discussed in \cite{matthieu2}.  In what follows we shall neglect it. This approximation corresponds to a real physical system where the soft spheres are replaced by point particles interacting with relaxed springs.   We now have:
\ba
\label{2}
\delta E= \frac{1}{2} \sum_{\langle  ij \rangle}[(\delta\vec{R_j}-\delta\vec{R_i}).\vec n_{ij}]^2  
\ea
If the system has too few contacts, there is  a set of displacements modes of vanishing restoring force and thus vanishing vibrational frequency.  These are the {\it soft modes}, or floppy modes. For these soft modes the energy $\delta E=0$ of Eq.~(\ref{2}) must vanish; therefore they must satisfy the $N_c$ constraint equations:
\be
\label{3}
(\delta\vec{R_i}-\delta\vec{R_j}).\vec n_{ij}=0 \ \hbox{ for all $N_c$ contacts}\ \langle ij \rangle
\ee
This linear equation defines the vector space of displacement fields that conserve the distances at first order between particles in contact. The particles can yield without restoring force if their displacements lie in this vector space.  Eq.~(\ref{3}) is purely geometrical and does not depend on the interaction potential. Each equation restricts the $dN$-dimensional space of $|\delta{\bf R}\rangle$ by one dimension.  Except when specific packing symmetries are present ( which is not the case here), these dimensions are independent, so that the number of independent soft modes is $dN - N_c$ \footnote[1]{Of these, $d(d+1)/2$ modes are dictated by the translational and rotational invariance of the energy function $\delta E[\delta {\bf R}]$. Apart from these, there are $dN - N_c - d(d+1)/2$ independent internal soft modes. In what follows we shall neglect the term $d(d+1)/2$, as it effect vanishes for large $N$.}.

\subsection{Isostaticity}

There are no internal soft modes in the rigid structure of a solid. This is true for a system of repulsive spheres, as soon as it  jams\footnote[2]{ The rattlers, particles without contacts, are removed in this argument, as they will lead to trivial zero translational modes.}.  Therefore jammed states must satisfy  $N_c \geq dN$, which is the Maxwell criterion for rigidity.  At the jamming transition, this inequality becomes an equality, as was verified in \cite{ohern}.  Such a system is called {\it isostatic}. The coordination number $z$ is then $z_c\equiv 2N_c/N \rightarrow 2d$. Physically the reason why $z=z_c$ rather than $z>z_c$ is the following:  if  $z>z_c$,  the system is over-constrained and contacts must be strained \cite{Tom1,moukarzel,roux}. This cannot be so at the jamming transition where the pressure and all contact forces vanish. Thus, the jamming transition is very different from rigidity percolation models where springs are deposited randomly on a lattice. In these models, which are in a sense at infinite temperature, there are both over-constrained and floppy regions at the rigidity threshold, and the percolating rigid cluster is a fractal object with dimension smaller than $d$.
In sphere packing the rigid system is a $d$-dimensional object. In  silica, we shall see that the spatial fluctuations of coordination are small, and we will argue that the vibrations of this system resembles those of sphere packing. 
In chalcogenide glasses where atoms of different valence are used, the properties of the covalent network near the rigidity threshold are still debated, but it has been proposed that isostaticity
 may also characterize the rigidity threshold \cite{boo}.    

\begin{figure}
\centering
\includegraphics[width=7cm]{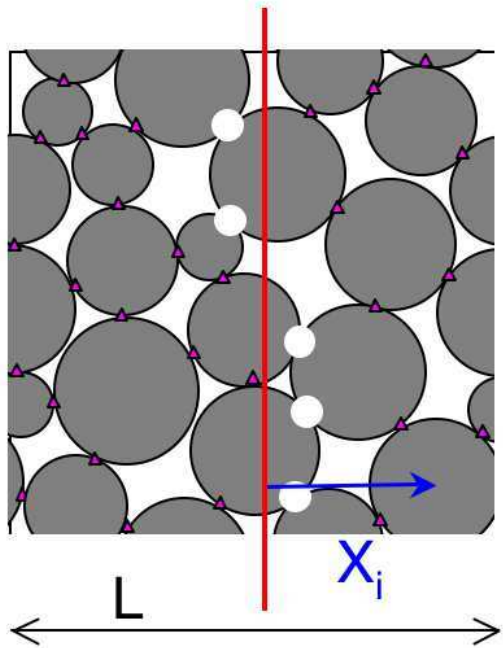}
\caption{(color online) Illustration of the boundary contact removal process described in the text.  Eighteen particles are confined in a square box of side $L$ periodically continued horizontally and vertically.  An isostatic packing requires 33 contacts in this two-dimensional system.  An arbitrarily drawn vertical line divides the system.  A contact is removed wherever the line separates the contact from the center of a particle.   Twenty-eight small triangles mark the intact contacts; removed contacts are shown by the five white circles.  }
\label{figs}
\end{figure}

\begin{figure}

\centering
\includegraphics[width=7cm]{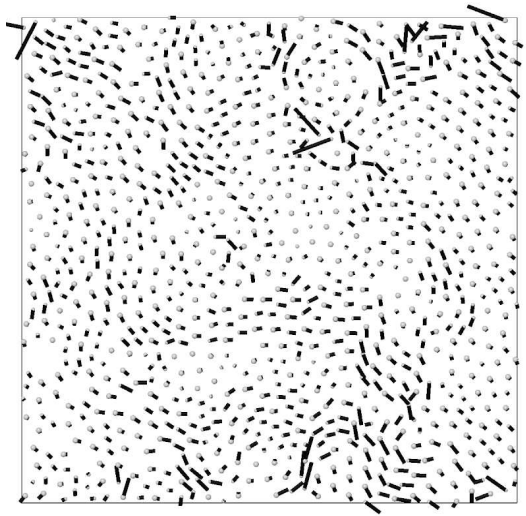}
\caption{ One soft mode in two dimensions for $N\approx 1000$ particles.  Each particle is represented by a dot. The relative displacement of the soft mode is represented by a line segment extending from the dot. The mode was created from a previously prepared isostatic configuration, periodic in both directions, following \cite{J}.  20 contacts along the vertical edges were then removed and the soft modes determined.  The mode pictured is an arbitrary linear combination of these
modes.}  
\label{softfig}
\end{figure}

An isostatic system is marginally stable: if $q$ contacts are cut, a
space of soft modes of dimension $q$ appears. For our coming argument
we need to discuss the extended character of these modes. In general
when only one contact $\langle ij \rangle$ is cut in an isostatic
system, the corresponding soft mode is not localized near $\langle ij
\rangle$. This arises from the non-locality of the isostatic condition
that gives rise to the soft modes, see \cite{these} for a theoretical derivation. 
When many contacts are severed, the extended character of the soft
modes that appear depends on the geometry of the region being cut. If
this region is compact many of the soft modes are localized.  For
example cutting all the contacts inside a sphere totally disconnects
each particle within the sphere. Most of the soft modes are then the
individual translations of these particles and are not extended
throughout the system. In what follows we will be particularly interested in the case where
the region of the cut is a hyperplane as illustrated in
Fig.~(\ref{figs}). In this situation occasionally particles in the
vicinity of the hyperplane can be left with less than $d$ contacts, so
that trivial localized soft modes can also appear. However we expect
that there is a non-vanishing fraction $q'$ of the total soft modes
that are not localized near the hyperplane, but rather extend over the
entire system, like the mode shown in Fig.~(\ref{softfig}). This assumption is checked numerically in \cite{matthieu2}. 
 We shall define extended modes more precisely in the next section.

\subsection{ $D(\omega)$ of an isostatic configuration}

We aim to show first that the density of states of an isostatic system does not vanish at zero frequency. $D(\omega)$ is the total number of modes per unit volume per unit frequency range.  Therefore we have to show that there are at least on the order of $\omega L^d$ normal modes with frequencies smaller than $\omega$ for any small $\omega$ in a system of linear size $L$. As we justify later, if proven in a system of size $L$ for $\omega\sim \omega_L\sim 1/L$, this property can be extended to a larger range of $\omega$ independent of $L$. Therefore it is sufficient to show that they are of the order of $L^{d-1}$ normal modes with frequency of the order of $1/L$, instead of the order of one such mode in a continuous solid. 
Our procedure for identifying the lowest frequency modes resembles that used for an ordinary solid.  An isolated block of solid has three soft modes that are simply translations along the three co-ordinate axes.  If the  block is enclosed in a rigid container, translation is no longer a soft mode.  However, one may recover the lowest-frequency, fundamental modes by making a smooth, sinusoidal distortion of the original soft modes.  We follow an analogous procedure to find the fundamental modes of our isostatic system.  First we identify the soft modes associated with the boundary constraints by removing these constraints.  Next we find a smooth, sinusoidal distortion of these modes that allows us to restore these constraints.
    
For concreteness we consider the three-dimensional cubical $N$-particle system $\cal S$ of Ref.~\cite{J} with periodic boundary conditions at the jamming threshold.  We label the axes of the cube by x, y, z.  $\cal S$ is isostatic, so that the removal of $n$  contacts allows exactly $n$ displacement modes with no restoring force. Consider for example the system $\cal S'$ built from $\cal S$ by removing the $q\sim L^2$ contacts crossing an arbitrary plane orthogonal to (ox); by convention at $x=0$, see Fig.~(\ref{figs}).  $\cal S'$, which has a free boundary condition instead of periodic ones along (ox), contains a space of soft modes of dimension $q$\footnote[3] { The balance of force can be satisfied in ${\cal S'}$ by imposing external forces on the free boundary. This adds a linear term in the energy expansion that does not affect the normal modes.}, instead of one such mode ---the translation of the whole system--- in a normal solid.  As stated above, we suppose that a subspace of dimension $q'\sim L^2$ of these soft modes contains only extended modes.  We define the {\it extension} of a mode relative to the cut hyperplane in terms of the amplitudes of the mode at distance $x$ from this hyperplane.  Specifically the extension $e$ of a normalized mode $|\bf \delta R \rangle$ is defined by $\sum_i\sin^2(\frac{x_i \pi}{L}) \langle i|{\bf \delta R} \rangle^2 =e$, where the notation $ \langle i|{\bf \delta R} \rangle$ indicates the displacement of the particle $i$ of the mode considered.  For example, a uniform mode with $ \langle i|{\bf \delta R} \rangle$  constant for all sites has $e=1/2$ independent of L. On the other hand, if $ \langle i|{\bf \delta R} \rangle=0$ except for a site $i$ adjacent to the cut hyperplane, the $x_i/L\sim L^{-1}$ and $e\sim L^{-2}$. We define the subspace of extended modes by setting a fixed threshold of extension $e_0$ of order 1 and thus including only soft modes $\beta$ for which $e_\beta>e_0$. As we discussed in the last section, we expect that a fixed fraction of the soft modes remain extended as the system becomes large. Thus if $q'$ is the dimension of the  extended modes vector space, we shall suppose that $q'/q$ remains finite as $L\rightarrow \infty$. The appendix presents our numerical evidence for this behavior.

We now use the vector space of dimension $q'\sim L^2$ of extended soft modes of $\cal{S'}$ to build $q'$ orthonormal trial modes of $\cal{S}$ of frequency of the order $1/L$.  Let us define $|\bf \delta R_\beta \rangle$ to be a normalized basis of this space, $1\leq \beta \leq q'$.  These modes are not soft in the jammed system $\cal S$ since they deform the previous $q$ contacts located near $x=0$.  Nevertheless a set of trial modes, $|\bf \delta R_\beta^* \rangle$, can still be formed by altering the soft modes so that they do not have an appreciable amplitude at the boundary where the contacts were severed.  We seek to alter the soft mode to minimize the distortion at the severed contacts while minimizing the distortion elsewhere. Accordingly, for each soft mode $\beta$ we define the corresponding trial-mode displacement $\langle i|{\bf \delta R}^* \rangle$ to be: 
\be
\label{rr}
\langle i|{\bf \delta R}_\beta^* \rangle \equiv C_\beta \sin(\frac{x_i \pi}{L}) \langle i|\bf \delta R_\beta \rangle
\ee
where the constants $C_\beta $  are introduced to normalize the modes.   $C_\beta $ depends of the spatial distribution of the mode $\beta$. If for example, a highly localized mode has $ \langle i|{\bf \delta R} \rangle=0$ except for a site $i$ adjacent to the cut plane, $C_\beta$ grows without bound as $L\rightarrow \infty$.  In the case of extended modes $C_\beta^{-2}\equiv \sum_{\langle ij \rangle} \sin^2(\frac{x_i \pi}{L})   \langle j|{\bf \delta R}_\beta \rangle^2= e_\beta > e_0$, and  therefore $C_\beta$ is bounded above by  $e_0^{-1/2}$.  The sine factor suppresses the problematic gaps and overlaps at the $q$ contacts near $x=0$ and $x=L$. The unit basis $|\bf \delta R_\beta \rangle$ can always be chosen such that the  $|\bf \delta R_\beta^* \rangle$ are orthogonal, simply because the modulation by a sine that relates the two sets is an invertible linear mapping in the subspace of extended modes.  Furthermore one readily verifies that the energy of each $|\bf \delta R_\beta^* \rangle$ is small, and that the sine modulation generates an energy of order $1/L^2$ as expected. Indeed we have from Eq.~(\ref{2}):
\be
\delta E =  C_\beta^2 \sum_{\langle ij \rangle} [(\sin(\frac{x_i \pi}{L}) \langle i|{\bf \delta R}_\beta \rangle- \sin(\frac{x_j \pi}{L}) \langle j|{\bf \delta R}_\beta \rangle) \cdot \vec n_{ij}]^2
\ee
Using  Eq.~(\ref{3}), and expanding the sine, one obtains:
\ba
\delta E \approx  C_\beta^2 \sum_{\langle ij \rangle} \cos^2(\frac{x_i \pi}{L}) \frac{\pi^2}{L^2} (\vec n_{ij} \cdot \vec e_x)^2 ( \langle j|{\bf \delta R}_\beta \rangle \cdot \vec n_{ij})^2 \\
\label{kk}
\leq  e_0^{-1}  (\pi/L)^2 \sum_{\langle ij \rangle}\langle j|{\bf \delta R}_\beta \rangle^2
\ea
where $\vec e_x$ is the unit vector along (ox), and where we used $|\cos| \leq 1$. The sum on the contacts can be written as a sum on all the particles since only one index is present in each term. Using the normalization of the mode $\beta$ and the fact that the coordination number of a sphere is bounded by a constant $z_{max}$ ($z_{max}=12$ for 3 dimensional spheres\footnote[12]{ In a polydisperse system $z_{max}$ could a priori be larger. Nevertheless Eq.~(\ref{kk}) is a sum on every contact where the displacement of only one of the two particles appears in each term of the sum. The corresponding particle can be chosen arbitrarily. It is convenient to choose the smallest particle of each contact. Thus when this sum on every contact is written as a sum on every particle to obtain Eq.~(\ref{kkk}), the constant $z_{max}$ still corresponds to the monodisperse case, as a particle cannot have more  contacts than that with particles larger than itself. }), one obtains:
\be
\label{kkk}
\delta E\leq   e_0^{-1}  (\pi/L)^2 z_{max} \equiv \omega_L^2
\ee
We have found on the order of $L^2$ trial orthonormal modes of frequency bounded by $\omega_L\sim 1/L$. This leads to $D(\omega_L)\sim L^2/\omega_L\sim L^0$, i.e. the average density of states is bounded below by a constant below frequencies of the order $\omega_L$.  This scaling argument is not entirely  rigorous as the orthonormal modes we use are not the normal modes. For completeness, we show how a variational procedure leads to a rigorous proof.   ${\cal M}$ is a positive symmetric matrix. Therefore if a normalized mode has an energy $\delta E$, we know that the lowest eigenmode has a frequency $\omega_0\leq \sqrt {\delta E}$. Such argument can be extended to a set of modes. If $m_\alpha$ is the $\alpha$'th lowest eigenvalue of ${\cal M}$ and if $e_\alpha$ is an orthonormal basis such that $\langle e_\alpha|{\cal M}| e_\alpha\rangle \equiv n_\alpha$ then the variational bound of A. Horn [Am. J. Math {\bf 76} 620 (1954)] shows that $\sum_1^q m_\alpha \leq \sum_1^q n_\alpha$.  Since $q n_q \geq \sum_1^q n_\alpha$, and since $\sum_1^q m_\alpha \geq \sum_{q/2}^q m_\alpha \geq (q/2) m_{q/2}$, we have $q m_p \geq (q/2) n_{q/2}$ as claimed. Thus, if there are $m$ {\it orthonormal} trial modes with energy $\delta E \leq \omega_t^2$, then there are at least $m/2$ eigenmodes with frequency smaller than $\sqrt 2\omega_t$. Therefore finding of the order of $L^{d-1}$ trial orthonormal modes with energy of order $ 1/L^2$ indeed leads to a non-vanishing density of normal modes.

 In what follows, the trial modes introduced in Eq.~(\ref{rr}), which are the soft modes modulated by a sine wave, shall be called ``anomalous modes'' to distinguish them from plane waves.  

To conclude, one may ask if this variational argument can be improved, for example by considering geometries of broken contacts different from the hyperplane surfaces we have considered so far.  When contacts are cut to create a vector space of extended soft modes, the soft modes must be modulated with a function that vanishes where the contacts are broken in order to obtain trial modes of low energy. On the one hand, cutting many contacts increases the number of trial modes. On the other hand, if too many contacts are broken, the modulating function must have many ``nodes'' where it vanishes. Consequently this function displays larger gradients and the energies of the trial modes increase. Cutting a surface (or many surfaces, as we shall discuss below) is the best compromise between these two opposit effects. Thus our argument gives a natural limit to the number of low-frequency states to be expected.

We may extend this argument to show that the bound on the average density of states extends to higher frequencies.  If the cubic simulation box were now divided into $m^3$ sub-cubes of size $L/m$, each sub-cube must have a density of states equal to the same $ D(\omega)$ as was derived above, but extending to frequencies on order of $m\omega_L$.  These subsystem modes must be present in the full system as well, therefore the  bound on $D(\omega)$ extends to $[0,m \omega_L]$.  We thus prove that the same bound on the average density of states holds down to sizes of the order of a few particles, corresponding to frequencies independent of $L$.  We note that in $d$ dimensions this argument may be repeated to yield a total number of modes, $L^{d-1}$, below a frequency $\omega_L \approx 1/L$, thus yielding a limiting nonzero density of states in any dimension.
We note that the trial modes of energy $\delta E\sim l^{-1}$ that we introduce by cutting the full system into subsystems of size $l$ are, by construction, localized to a distance scale $l$. Nevertheless we expect that these trial modes will hybridize with the trial modes of other, neighboring, subsystems; the corresponding normal modes will therefore not to be localized to such short length scales.       
\begin{figure}
\centering
\includegraphics[angle=0,width=7cm]{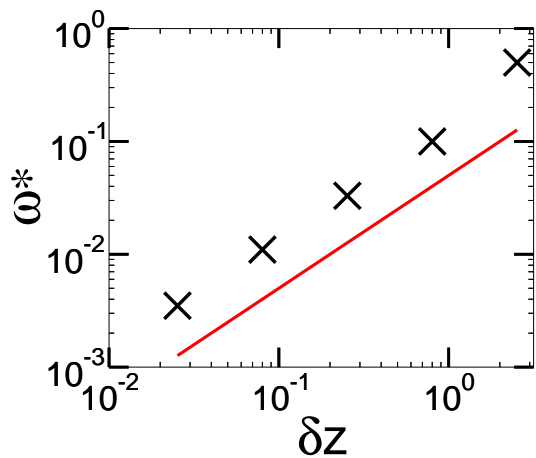}
\caption{(color online) Scaling of $\omega^*$ with the excess coordination number, $\delta z$ in the system with relaxed springs. The line has a slope one.}
\label{sans}
\end{figure}

\subsection{$D(\omega)$ when $\delta z>0$ }

When the system is compressed and moves away from the jamming transition, the simulations show that the extra-coordination number  $\delta z \equiv z-z_c$ increases. This causes $\Delta N_c=N \delta z/2 \sim L^d \delta z$ extra constraints to appear in Eq.~(\ref{2}).  Cutting the boundaries of the system, as we did above, relaxes $q\sim L^{d-1}$ constraints.  For a large system, $L^d\delta z> L^{d-1}$ and thus $q<\Delta N_c$.  Thus Eq.~(\ref{2}) is still over-constrained and there will be no soft modes in the system.  
However, as the systems become smaller, the excess number of constgraints, $\Delta N_c$, diminishes; for $L$ smaller than some $l^*\sim \delta z^{-1}$, $q$ becomes larger than $\Delta N_c$the system is again under-constrained as was already noticed in \cite{Tom1}.  This allows one to build low-frequency modes in subsystems smaller than $l^*$.  These modes appear above a cut-off frequency
$\omega^*\sim l^*{}^{-1}$; they are the ``anomalous modes" that contribute to the flat plateau in
$D(\omega)$ above
$\omega^*$.  
In other words, anomalous modes with characteristic length
smaller than
$l^*$ are not affected very much by the extra contacts, and the density of states is
unperturbed above a frequency $\omega^* \sim \delta z$.
This scaling is checked numerically in  Fig.~\ref{sans}. This prediction is in very good
agreement with the data up to $\delta z\approx 2$. 
 
At frequencies lower than $\omega^*$ we expect the system to behave as a disordered, but not poorly-connected, elastic medium.  These vibrational modes are  similar to the plane waves of a continuous elastic body.  We refer to these modes as ``acoustic modes''.  
The behavior of such systems near the jamming threshold thus depend on the frequency $\omega$ at which they are observed. For $\omega>\omega^*$ the system behaves as an isostatic system, and for $\omega<\omega^*$ it behaves as a continuous elastic medium.  Equivalently $l^*$ is the distance below which a continuous elastic  description is not a good approximation. This was confirmed numerically in \cite{wouter} where it was found that $l^*$ characterize the stress heterogeneities after a point force is imposed. 


\section{The case of Amorphous silica}

\begin{figure}
\centering
\includegraphics[angle=270,width=7cm]{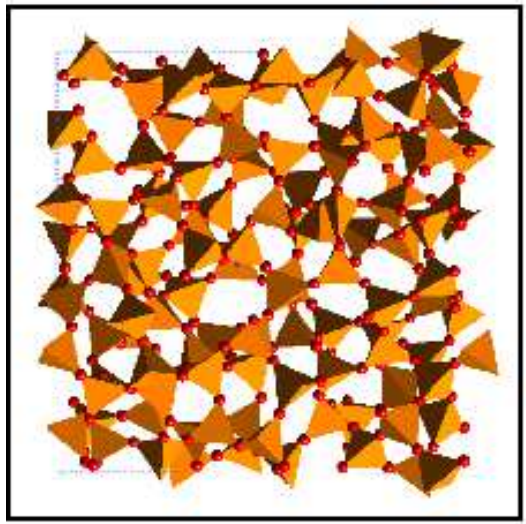}
\caption{Rigid unit modes model applied to silica. Trachenko et al.\cite{kostya}.}
\label{sd}
\end{figure}
\begin{figure}
\centering
\includegraphics[angle=0,width=7cm]{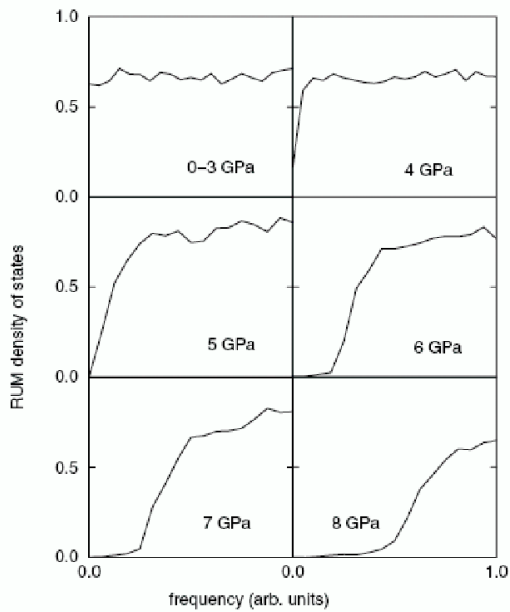}
\caption{Density of  rigid unit modes for silicate at different pressure. Trachenko et al.\cite{dove}. }
\label{dd}
\end{figure}

 
 Silica is perhaps the most common glass. It is also known to have one of the strongest  excess of low-frequency modes, or boson peak, see \cite{nakayama}  for a review of empirical results and models. In this paragraph we propose an explanation for its density of states at low-frequency, and for the nature of the excess-modes.   In this glass (or more generally aluminosilicates) the forces within the tetrahedra $SiO_4$  are much stronger than the forces that act between them \cite{hammonds}: it is easier to rotate two linked tetrahedra, that is to bend the Si-O-Si angle,  than to distort one tetrahedron:  the bending energy of Si-o-Si is roughly 10 times smaller than the stretching of the contact Si-o \cite{ff}. This suggests to model such glass as an assembly of linked tetrahedra loosely connected at corners: this is the ``rigid unit modes'' model \cite{heine}.  In this model the tetrahedra are characterized by a unique parameter, a stiffness $k$ \footnote[9]{In fact the rigidity of a tetrahedron induced by the covalent bonds should be characterized by 3 parameters corresponding to different deformations of the tetrahedron. If these parameters are of similar magnitude, as one expects for example for silica,  this does not change qualitatively the results discussed here.}. Recently this model was used to study the vibrations of silica  \cite{dove}.  The authors first generate realistic configurations of $SiO_2$ at different pressures using molecular dynamics simulations. At low pressure, they obtain a perfect tetrahedral network. When the pressure becomes large, the coordination of the system increases with the formation of 5-fold defects. Once these microscopic configurations are obtained, the rigid unit model is used and the system is modeled as an assembly of elastic tetrahedra, see Fig.(\ref{sd}). Then, the density of states of such network is computed. The results are shown in Fig.(\ref{dd}). One can note the obvious similarity with the density of states near jamming of Fig.(\ref{f1}). We argue that the cause is identical, and that the excess-modes correspond to the anomalous modes made from the soft modes, rather than to one-dimensional modes as proposed in \cite{dove}. Indeed, a tetrahedral network is isostatic, see e.g. \cite{kostya}. The counting of degrees of freedom goes as follows: on the one hand each tetrahedron has 6 degrees of freedom (3 rotations and 3 translations). On the other hand, the 4 corners of a tetrahedron bring each 3 constraints shared by 2 tetrahedra, leading to 6 constraints per tetrahedron. Thus the system is isostatic. When the pressure increases the coordination increases too, leading to the erosion of the plateau in the density of states discussed earlier. 

\begin{figure}
\centering
\includegraphics[angle=0,width=13cm]{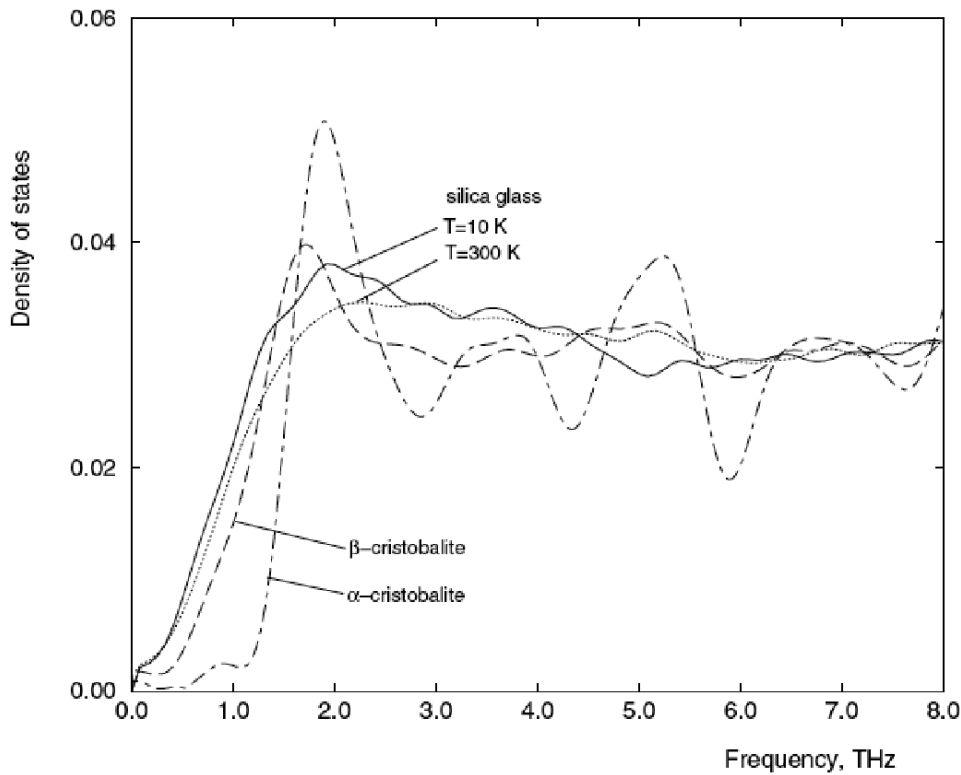}
\caption{Density of states of silica glass (at temperatures of 10 and 300 Kelvins), $\alpha$-cristobalite and $\beta$-cristobalite. This figure is taken from the simulations of \cite{kostya}.  }
\label{cry}
\end{figure}

These predictions fail to describe the spectrum of silica at low frequencies, where the weak interactions, in particular the bending of the Si-O-Si bond and the  Van der Waals interactions cannot be neglected.
The main effect of these interactions is to shift the spectrum  of anomalous modes by some frequency scale $\zeta$. To estimate $\zeta$ we use the stiffness of the Si-O-Si bending interaction obtained {\it ab initio} \cite{ff}, and the molecular mass to form a frequency. One finds $\zeta\approx 1.4 Thz$. We therefore expect the spectrum  of silica to display a plateau above a frequency of order 1 Thz. This is indeed what is observed in simulations: silica glass present a well-defined plateau above the boson peak frequency, as appears in the numerical results shown in Fig.(\ref{cry}).  

Our argument does not involve disorder. Thus it must also apply for the crystals of the same composition and similar densities such as the $\alpha$ and $\beta$-cristobalite, since these crystalline structures are formed, as silica,  by  SiO$_4$ tetrahedra connected at the corners.  $\beta$-cristobalite has the structure of the diamond, in which the tetrahedra correspond to the 4 carbons bonded  to a central carbon, whereas  $\alpha$-cristobalite has a tetragonal structure. Empirically a boson peak is observed in all these materials \cite{orenbarre}. Numerically, a plateau indeed appears in $D(\omega)$ at roughly the same frequency in the cristobalite $\alpha$ and $\beta$ \cite{kostya} and in the glass, as shown in Fig.(\ref{cry}).  In the crystalline case the plateau corresponds to a Van Hove singularity and to the accumulation of optical bands. More generally, there are other crystals showing an excess  density of states at frequencies of the order of the boson peak frequency of the corresponding glass \cite{nakayama,leadbetter,caplin,bilir}. According to the present argument this should be the case as long as the crystal and the amorphous structure share a similar connectivity (which is not the case, for example, for sphere packing).
 
Although disorder is not relevant to compute the density of states, it affects the nature of the vibrational modes.  Transport is very different in  silica glass and in cristobalite. Thus the peculiarity of the amorphous state  lies in the {\it nature} of the excess-modes,  not in the density of states \cite{nakayama}. It is useful to note the  parallel between cristobalite and  silica glass on the one hand and cubic lattice and the jamming threshold of elastic spheres on the other hand.  In both cases the amorphous solid and the crystal have a similar density of states, but the anomalous modes in the amorphous phase are not plane waves. Disorder strongly affects the anomalous modes, and makes them very heterogeneous as shown in Fig(\ref{softfig}). 

In chalcogenide glasses, assuming weak fluctuations of coordination (which may be a naive assumption in that case)  our argument leads to a Boson Peak frequency increasing for  over-constrained networks as $\omega^*\sim (\zeta^2 +a (x-x_c)^2)^{1/2}$, where $x_c$ is the critical composition above which floppy modes vanish,  $a$ is a numerical constant and $\zeta$ is the finite frequency of the floppy modes for $x<x_c$, induced by van der Waals interactions.  When the coordination becomes too large the Boson Peak vanishes, as occurs for example for amorphous silicon. 

\section{Conclusion}

There exists several analytical theories of the Boson Peak based on specific models of the disorder. Some of the authors have 
investigated models in which the disorder enters via local defects \cite{local}; intrinsic disorder in the harmonic force constants, see e.g.  \cite{schi,tu}, in which the boson peak frequency marks a crossover from acoustic-like sound excitations to a  disorder-dominated regime, or infinite temperature systems where particle positions are random \cite{parisi2,parisi3,mezard}. These models are able to generate a Boson Peak, to explain the presence of a plateau in the thermal conductivity of glasses, see e.g.  \cite{schi}, and can describe sound attenuation at those frequencies \cite{att}.  With respect to those approaches, the rigidity, or coordination -based  description has the  advantages of: (i) relating the boson peak  to a microscopic observable, the coordination number, which can be inferred from the composition of the glass for covalent networks, and more generally from the knowledge of the microscopic  structure (ii) explaining why crystalline structures can also present a peak and allows to compute its position (iii)  introducing a length scale $l^*$ characterizing the modes responsible for the peak. This length scale was shown to  describe the extension of the force chains \cite{wouter}.  These results can be extended to colloidal glasses \cite{brito} and systems with rapidly decaying long-range potentials, such as a Lennard-Jones \cite{these,ning}. 

\section{Acknowledgment}

It is a pleasure to thanks C. Brito, J-P Bouchaud, A. Liu, S. Nagel, L. Silbert, V. Vitelli, T. Witten  and N. Xu for  stimulating discussions.

\end{document}